\documentclass[twocolumn,showpacs,prb,superscriptaddress,floatfix]{revtex4}

\usepackage{graphicx}
\usepackage{amsmath}

\newcommand{\figurewidth}{\columnwidth}

\newcommand{\av}{{\mathrm{av}}}

\newcommand{\sub}[1]{\ensuremath{_\textrm{\scriptsize #1}}}

\begin{document}

\title{Critical behavior of the three- and ten-state short-range Potts
glass:\\A Monte Carlo study}

\author{L.~W.~Lee}
\affiliation{Department of Physics,
University of California,
Santa Cruz, California 95064, USA}

\author{H.~G.~Katzgraber}
\affiliation{Theoretische Physik,
ETH Z\"urich, CH-8093
Z\"urich, Switzerland}

\author{A.~P.~Young}
\email[Email: ]{peter@physics.ucsc.edu}
\affiliation{Department of Physics,
University of California,
Santa Cruz, California 95064, USA}

\date{\today}

\begin{abstract}
We study the critical behavior of the short-range $p$-state Potts spin glass
in three and four dimensions using Monte Carlo simulations. In three
dimensions, for $p = 3$, a finite-size scaling analysis of the correlation
length shows clear evidence of a transition to a spin-glass phase at
$T_{\rm c} \simeq 0.273$ for a Gaussian distribution of interactions and 
$T_{\rm c} \simeq
0.377$ for a $\pm J$ (bimodal) distribution.  These results indicate that the
lower critical dimension of the three-state Potts glass is below three. By
contrast, the correlation length of the ten-state ($p = 10$) Potts glass in
three dimensions remains small even at very low temperatures and thus shows no
sign of a transition. In four dimensions we find that the $p = 3$ Potts glass
with Gaussian interactions has a spin-glass transition at $T_{\rm c} \simeq 0.536$.
\end{abstract}

\maketitle

\section{Introduction}
\label{sec:introduction}

There are two main motivation for studying the Potts glass: First,
the model for which the number of Potts states $p$ is equal to $3$ is
a model for orientational glasses\cite{binder:92}
(also known as quadrupolar glasses), which are randomly diluted molecular
crystals where quadrupolar moments freeze in random orientations upon lowering
the temperature due to randomness and competing interactions. This is similar
to spin glasses where spins are frozen in random directions in the spin-glass
phase. However, unlike conventional spin-glass systems\cite{binder:86} 
with Ising or vector spins, the Potts model does not have
spin inversion symmetry, except for the special case with
$p = 2$ in which case the Potts model reduces to the Ising model.
In mean-field theory (i.e., for the
infinite-range version of the model) it is found\cite{gross:85}
that the three-state Potts glass has a
finite transition temperature to a spin-glass-like phase.

Second, the Potts glass could help us
understand the structural glass transition of supercooled fluids.  In the 
mean-field (i.e., infinite-range) limit, the Potts glass with $p > 4$ has 
a behavior quite different from that of the Ising spin glass
since \textit{two}\cite{kirkpatrick:87,kirkpatrick:88} transitions occur 
in the Potts case. As temperature is lowered, the first transition
is a dynamical transition at $T_{\rm d}$ below which the autocorrelation
functions of the Potts spins do not decay to zero in the long-time limit.
The second transition is a static transition at $T_{\rm c} < T_{\rm d}$ 
where a static order parameter, described by ``one-step replica symmetry 
breaking,'' appears discontinuously. At the mean-field level, it turns 
out that the the equations which describe the dynamics of the $p$-spin 
Potts glass for $T > T_{\rm d}$ are almost identical to the
the mode-coupling equations; see Ref.~\onlinecite{goetze:92} and references
therein, which describe structural glasses as they are cooled
towards the glass transition. Hence Potts glasses with $p > 4$
and structural glasses seem to be connected, at least in mean-field theory.

However, it is important to understand to
what extent these predictions for phase transitions in the infinite-range Potts
glass are also valid for more realistic \textit{short-range} 
models. Here we consider this question by performing Monte Carlo
simulations on the three-state and ten-state Potts glass models.

Earlier work on short-range three-state Potts glasses in three dimensions have used
zero-temperature domain-wall renormalization group
methods,\cite{banavar:89a,banavar:89b} Monte Carlo
simulations,\cite{scheucher:90,scheucher:92,reuhl:98}
and high temperature series expansions.\cite{singh:91,schreider:95,lobe:99}
The results of the domain-wall and Monte Carlo studies have generally
indicated
that the lower critical dimension $d_l$ is equal to, or close to, $3$. The
series expansion results tend to be better behaved in higher dimensions and
less reliable in $d = 3$, but Schreider and Reger\cite{schreider:95} have
also argued in favor of $d_l = 3$ from an analysis of their series. 
Overall, though, it seems to us that conclusive evidence for the value of the
lower critical dimension is still lacking. In this
paper we therefore investigate the lower critical dimension of the
three-state Potts glass by Monte Carlo simulations of this model in
three space
dimensions using the method that has been most successful in elucidating the
phase transition in Ising\cite{ballesteros:00}
as well as vector\cite{lee:03} spin glasses: a
finite-size scaling analysis of the finite-size correlation length.

In four space dimensions, Scheucher and Reger\cite{scheucher:93} performed Monte
Carlo simulations, finding that the critical temperature $T_{\rm c}$ is finite 
and a value for the
order parameter exponent $\beta$ close to zero, as well as a correlation
length exponent $\nu$ consistent with $1/2$. These are
the expected\cite{gross:85} values
at the discontinuous transition in the mean-field Potts
model for $p > 4$. It is surprising that these should be found in four
dimensions, since, as for Ising and vector spin glasses,\cite{harris:76}
the upper critical dimension is $6$. To try
to clarify this situation we therefore also apply the finite-size scaling
analysis of the correlation length to the three-state Potts glass in $d=4$.

Earlier work on the ten-state Potts glass in three dimensions by
Brangian {\it et al.}\cite{brangian:02b,brangian:03}
found that, in contrast to
the mean-field version of the same model, \textit{both}
the static and dynamic
transitions are wiped out. Relaxation times follow simple
Arrhenius behavior, and judging by the lack of size dependence in the
results, the correlation length is presumed to be small, although it was not
calculated explicitly. In this paper we calculate the correlation length
and find that it is always less than one lattice spacing, consistent with the
results of Brangian {\it et al}.

In Sec.~\ref{sec:model}
we discuss the model and the numerical method that we have used.
We show the results of the simulations for the three-state model in
Sec.~\ref{sec:results_3state} and for the ten-state model in
Sec.~\ref{sec:results_10state}.
Section \ref{sec:conclusions} contains a summary.

\section{Model and method}
\label{sec:model}
In the Potts glass, the Potts spin on each site $i$ can
be in one of $p$ different states, $n_i = 1, 2, \ldots, p$.
If neighboring spins at $i$ and $j$ are in the same state, the energy is
$-J_{ij}$, whereas if they are in different states,
the energy is zero. Thus the Hamiltonian of the model is given by
\begin{equation}
{\mathcal H} = -\sum_{\langle i, j \rangle} J_{ij} \delta_{{n_i} {n_j}} \, , 
\label{ham}
\end{equation}
in which the summation is over nearest-neighbor pairs. The sites lie on a
hypercubic lattice with $N = L^d$ sites, and we consider dimensions $d=3$ 
and $4$. The interactions $J_{ij}$
are independent random variables with either
a Gaussian distribution with mean $J_0$ and standard deviation unity,
\begin{equation}
{\mathcal P}(J_{ij}) = \frac{1}{\sqrt{2\pi}} e^{-(J_{ij} - J_0)^2/2},
\label{eq:gau}
\end{equation}
or a bimodal ($\pm J$) distribution with zero mean,
\begin{equation}
{\cal P}(J_{ij}) = \frac{1}{2}\left[\delta(J_{ij}-1) + \delta(J_{ij}+1)\right].
\end{equation}

It is convenient to rewrite the Potts glass Hamiltonian using the ``simplex''
representation where the $p$ states are mapped to the $p$ corners of a
hypertetrahedron in $(p - 1)$-dimensional space. The state at each site is
represented by
a $(p - 1)$-dimensional unit vector $\mathbf{S}_i$, which takes one of $p$
values, $\mathbf{S}^\lambda$ satisfying
\begin{equation}
\mathbf{S}^{\lambda} \cdot \mathbf{S}^{\lambda'} =
\frac{p\;\delta_{\lambda\lambda'} - 1}{p-1}
\label{simplex}
\end{equation}
($\lambda = 1, 2, \cdots, p$).
In the simplex representation, the Potts Hamiltonian is similar to 
a $(p-1)$-component vector spin-glass model,
\begin{equation}
{\mathcal H} = -\sum_{\langle i, j \rangle} J'_{ij} \mathbf{S}_i \cdot
\mathbf{S}_j ,
\label{simplexham}
\end{equation}
ignoring an additive constant, where 
\begin{equation}
J'_{ij} = \frac{(p-1)}{p}J_{ij} .
\label{J'}
\end{equation}
However, an important difference compared with a vector spin glass is that,
except for $p=2$, if $\mathbf{S}^{\lambda}$ is one of the (discrete) allowed
vectors, then its inverse $-\mathbf{S}^{\lambda}$ is \textit{not} 
allowed.

In mean-field theory, assuming the transition is continuous,
the spin-glass transition temperature is given
by\cite{gross:85}
\begin{equation}
T_{\rm c}^{\rm MF} = \frac{z^{1/2}}{p} [(J_{ij}-J_0)^2]_\av^{1/2} 
                   = \frac{z^{1/2}}{p} 
	           = \frac{z^{1/2}}{p - 1} [(J'_{ij}-J_0')^2]_\av^{1/2} ,
\label{TcMF}
\end{equation}
where $z = 2d$ is the number of nearest neighbors and $[ \cdots ]_\av$
denotes an average over disorder.
Actually, in mean-field theory
there is a discontinuity in the order parameter for $p > 4$ and
the transition then occurs at
a higher temperature.\cite{gross:85} For the case of $p=10$ this is found to
be at\cite{brangian:02c} $T_{\rm c} = 0.28$, rather than Eq.~(\ref{TcMF}),
which predicts $T_{\rm c} = 0.24$.
Hence, the mean-field transition temperatures for the models discussed in this
paper are
\begin{equation}
T_{\rm c}^{\rm MF} = \left\{
\begin{array}{lll}
0.82 &\;&\mbox{three-state},\; d = 3, \\
0.94 &\;&\mbox{three-state},\; d = 4, \\
0.28 &\;&\mbox{ten-state},\; d = 3.
\end{array}
\right.
\label{TSGMF}
\end{equation}

The spin-glass order parameter for a Potts spin glass with wave 
vector ${\bf k}$, $q^{\mu\nu}({\bf k})$, is defined to be
\begin{equation}
q^{\mu\nu}({\bf k}) = \frac{1}{N} \sum_i S_i^{\mu(1)} S_i^{\nu(2)}
e^{i {\bf k} \cdot {\bf R}_i},
\end{equation}
where $\mu$ and $\nu$ are components of the spin in the simplex
representation and ``$(1)$'' and ``$(2)$'' denote two
identical copies (replicas) of the system with the same disorder.  
The wave-vector-dependent
spin-glass susceptibility $\chi_{\rm SG}({\bf k})$ is then given by
\begin{equation}
\chi_{\rm SG}({\bf k}) = N \sum_{\mu,\nu} [\langle \left|q^{\mu\nu}({\bf
k})\right|^2 \rangle ]_\av ,
\end{equation}
where $\langle \cdots \rangle$ denotes a thermal average.

The correlation length, $\xi_L$, of a system of size $L$ is
defined\cite{ballesteros:00} by
\begin{equation}
\xi_L = \frac{1}{2 \sin (k_\mathrm{min}/2)}
\left[\frac{\chi_{\rm SG}({\bf 0})}{\chi_{\rm SG}({\bf k}_\mathrm{min})} 
- 1\right]^{1/2},
\end{equation}
where ${\bf k}_\mathrm{min} = (2\pi/L, 0, 0)$ is the smallest nonzero
wave vector.
Since the ratio of $\xi_L/L$ is dimensionless, it satisfies the finite-size 
scaling form
\begin{equation}
\frac{\xi_L}{L} = \widetilde{X}\left(L^{1/\nu}[T-T_{\rm c}]\right),
\label{eq:fss}
\end{equation}
where $\nu$ is the correlation length exponent, $T_{\rm c}$ is the
transition temperature, and $\widetilde{X}$ a scaling
function. At the transition temperature,
$\xi_L/L$ is independent of $L$, and so this quantity is particularly
convenient for locating the transition. Once $T_{\rm c}$ has been obtained from
the intersection point, one determines the exponent $\nu$ by requiring that
the data collapse onto a single curve when plotted as a function of
$L^{1/\nu} (T - T_{\rm c})$.

We are also interested
in obtaining the critical exponent $\eta$ which characterizes the power-law
decay of the correlation function at criticality. The two exponents
$\nu$ and $\eta$
completely characterize the critical behavior, since the other 
exponents can be obtained from them using scaling relations.\cite{yeomans:92}
From scaling theory, we expect the spin-glass susceptibility
$\chi_{\rm SG}$ to vary as
\begin{equation}
\chi_{\rm SG} = L^{(2-\eta)}\widetilde{C}\left(L^{1/\nu}[T-T_{\rm c}]\right).
\label{eq:chisg_fss}
\end{equation}

To perform the scaling fits systematically, we use the method recently
proposed in Ref.~\onlinecite{katzgraber:06}.
By assuming that the scaling function can be approximated in the vicinity of
the critical temperature by a third-order polynomial
(with coefficients $c_i$, $i$=1--4), a nonlinear
fit can be performed on the data for $\xi_L$ with six parameters
($c_i$, $T_{\rm c}$, and $\nu$) using Eq.~(\ref{eq:fss}) in order to
determine the optimal finite-size scaling.
For the $\chi_{\rm SG}$ data, we have an additional parameter $\eta$ from
Eq.~(\ref{eq:chisg_fss}), so the fit involves 7 parameters. It is necessary
to choose a temperature range within the scaling region to do the fits.
We limit our temperature range to where results from scaling the data
in $(T-T_{\rm c})$ and $(\beta-\beta_{\rm c})$, where $\beta = 1/T$ is
the inverse temperature, agree within error bars and quote
the results obtained from the fits with $(T-T_{\rm c})$.
To estimate the errors we used a bootstrap procedure, described in
Ref.~\onlinecite{katzgraber:06}. If there are $N_{\rm samp}$ samples for a
given size, we
generate $N_{\rm boot} = 100$
random sets of the samples (with replacement) with
$N_{\rm samp}$ samples in each set, do the
analysis for each set, and take the standard deviation of the results from the
sets to be the error.
We emphasize that this gives the \textit{statistical} errors only. In
addition, there can be \textit{systematic} errors, from corrections to
finite-size scaling which are difficult to estimate from simulations on only a
modest range of sizes, as is possible here.

Spin-glass simulations are hampered by long equilibration times, so
we use parallel tempering,\cite{hukushima:96,marinari:98b}
which has proven useful in speeding up equilibration.
In this approach, copies
of the system with the same interactions are simulated at several
temperatures. To check for
equilibration, we use the following relation which holds if the couplings
$J_{ij}$ are drawn from a Gaussian distribution with mean $J_0$ and unit
standard deviation:
\begin{equation}
\frac{2 T [U]_\av}{z} = \left[\langle \delta_{n_i n_j} \rangle^2\right]_\av -
		  \left[\langle\, (\delta_{n_i n_j})^2 \,\rangle \right]_\av
			- J_0 T \left[\langle \delta_{n_i n_j}
			  \rangle\right]_\av,
\label{eq:eqil}
\end{equation}
where $U = (1/N)\langle {\mathcal H}\rangle$ is the energy per spin for 
a given disorder realization.
Equation (\ref{eq:eqil}) is obtained\cite{katzgraber:01} by integrating by parts
the expression for the energy with respect to the interactions $J_{ij}$.
Note that the square can be omitted in the second term on the right of
Eq.~(\ref{eq:eqil}) (since $\delta_{n_i n_j}$ only takes values 1 and 0).
The first term on the right is calculated from two copies at the same
temperature---i.e.,
$\left[\langle \delta_{n_i n_j}\rangle^{(1)}
\langle\delta_{n_i n_j} \rangle^{(2)}\right]_\av$.
Hence, if $N_T$ is the number of
temperatures, the total number of copies simulated
(with the same interactions) is $2 N_T$.  As the simulation proceeds,
the left-hand side (LHS) of Eq.~(\ref{eq:eqil}) approaches the equilibrium value
from above, while
the right-hand side (RHS) approaches the (same) equilibrium value from below
since the spins in the two copies are initially uncorrelated.
When both quantities are simultaneously plotted as a
function of Monte Carlo sweeps, the two curves merge when equilibration
is reached and remain together subsequently.\cite{katzgraber:01}
Hence requiring that Eq.~(\ref{eq:eqil}) is satisfied
is a useful test for equilibration.

\begin{center}
\begin{figure}
\includegraphics[width=\figurewidth]{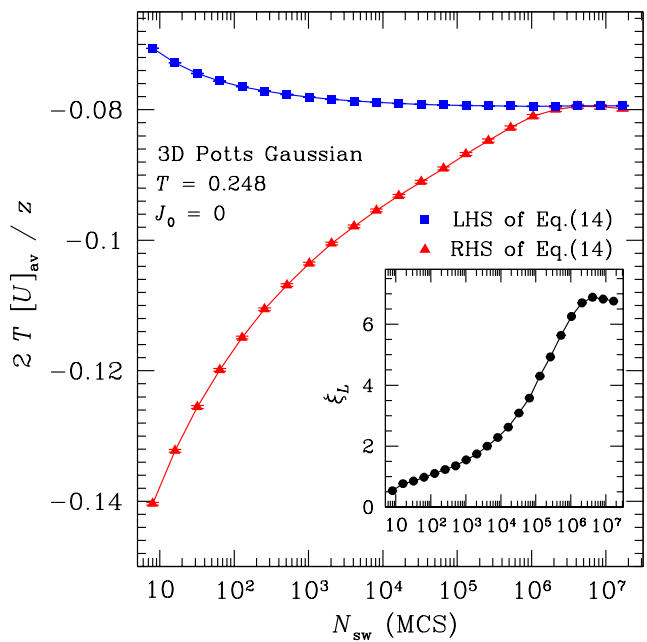}
\caption{(Color online) An equilibration plot showing
the left-hand side (upper curve) and right-hand side (lower curve)
of Eq.~(\ref{eq:eqil}). The data are for
the three-state Potts glass in $d=3$ for $L=12$
at $T=0.248$ as a function of equilibration time measured in Monte Carlo
sweeps, $N_{\rm sw}$. An equal number of sweeps are also performed for
measurement, so the total number of sweeps is $2 N_{\rm sw}$,
respectively. The two sets of data merge at long times, as expected, and then
remain independent of time.
The inset shows the correlation length $\xi_L$ for
the corresponding number of Monte Carlo sweeps.
}
\label{fig:equi_plot}
\end{figure}
\end{center}

An example of the equilibration test is shown in
Fig.~\ref{fig:equi_plot} for the three-state Potts glass in $d=3$ for $L=12$
at $T=0.248$. In the main panel, the upper curve is the
energy term on the left-hand side of Eq.~(\ref{eq:eqil})
while the lower curve corresponds to the right-hand side of
Eq.~(\ref{eq:eqil}) as a function of equilibration time measured in
Monte Carlo sweeps (MCS).
Note that one sweep consists of one Metropolis
sweep of single spin-flip moves on each copy,
followed by a sweep of global moves in which spin
configurations at neighboring temperatures are swapped.
Figure \ref{fig:equi_plot} shows the
number of sweeps used for measurements, $N_{\sub{sw}}$; in addition, an
equal number of sweeps have been used for equilibration, so the total number
of sweeps is $2 N_{\sub{sw}}$.
We see that the two curves merge in this case at around $10^7$ MCS and
stay within error bars after that, indicating that equilibration has been
reached. The inset of Fig.~\ref{fig:equi_plot}
shows the correlation length $\xi_L$ as a function of Monte Carlo sweeps.
The data increase until
$\sim 10^7$ MCS where they saturate. This indicates that the
equilibration time for the correlation length is in agreement with 
the equilibration time determined from the requirement that the two
sides of Eq.~(\ref{eq:eqil}) agree.

Unfortunately, the equilibration test using Eq.~(\ref{eq:eqil}) only works
for a Gaussian bond distribution. For the bimodal disorder distribution,
we study how the results vary when the
simulation time is successively increased by factors of 2 (logarithmic
binning of the data). We require that the last three measurements for all 
observables agree within error bars.

For the Potts glass with $p > 2$, an additional issue arises, which is 
not present for the Ising ($p = 2$) case:
At low temperatures ferromagnetic correlations develop\cite{elderfield:83} 
due to lack of spin-inversion symmetry, even for a symmetric 
disorder distribution ($J_0 = 0)$.
Since we want to study the glassy phase, rather than the ferromagnetic
phase,
we can mitigate this undesirable feature by choosing a negative value for
$J_0$. A choice of $J_0=-1$
seems sufficient for the ten-state Potts glass,\cite{brangian:03}
while for the three-state Potts glass, the ferromagnetic correlations are
small, as we shall show, so we set $J_0 = 0$.

\section{Three-state Potts glass}
\label{sec:results_3state}

\subsection{Gaussian disorder ($\boldsymbol{d=3}$)}

\begin{center}
\begin{figure}
\includegraphics[width=\figurewidth]{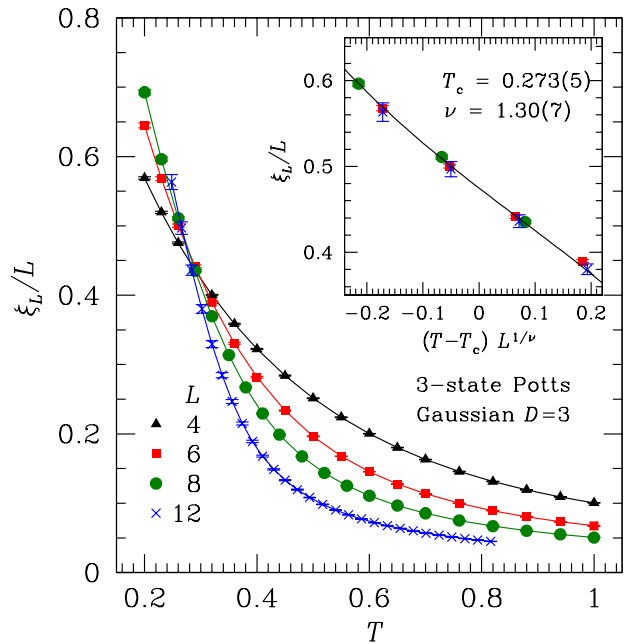}
\caption{(Color online) 
Finite-size correlation length $\xi_L/L$ vs $T$ for the three-state Potts
glass with Gaussian couplings
in three dimensions. The curves intersect at $T_{\rm c} \simeq 0.273$.
The inset shows a scaling plot according to Eq.~(\ref{eq:fss}) using
$T_{\rm c} = 0.273$ and $\nu=1.30$. The solid line is the third-order 
polynomial used in the fit.
}
\label{fig:xi_L_3state_gau_D3}
\end{figure}
\end{center}

We first discuss results for the three-state Potts glass with Gaussian
interactions. 
The simulation parameters are shown in Table.~\ref{table:params_3state_Gau_3D}.
As shown in Fig.~\ref{fig:xi_L_3state_gau_D3}, the data for $\xi_L/L$ 
intersect at $T \simeq 0.27$.
The common intersection provides good evidence
for a spin-glass transition at this temperature. It would be
desirable to obtain data in this temperature range for larger sizes, but 
unfortunately equilibration times are prohibitively long. Interestingly,
the same transition temperature $T_{\rm c} \simeq 0.27$ was proposed earlier
by Banavar and Cieplak\cite{banavar:89a,banavar:89b} who studied the
free-energy sensitivity to changes in boundary conditions at finite
temperature using very small system sizes $L=2$ and 3.

To determine $T_{\rm c}$ and $\nu$, we
analyze the $\xi_L$ data for $L\ge 6$ using the method discussed
in Sec.~\ref{sec:model} for the temperature range $0.20 < T < 0.32$
for all $L$ and obtain
\begin{equation}
T_{\rm c} = 0.273(5), \qquad \nu = 1.30(7)\, .
\label{res_3d_gauss}
\end{equation}
A finite-size scaling analysis of $\xi_L/L$ according to Eq.~(\ref{eq:fss})
is shown in the inset of Fig.~\ref{fig:xi_L_3state_gau_D3} for $L\ge 6$.
The solid line represents the third-order polynomial that we have used to
approximate the scaling function. The data collapse well.

\begin{center}
\begin{figure}
\includegraphics[width=\figurewidth]{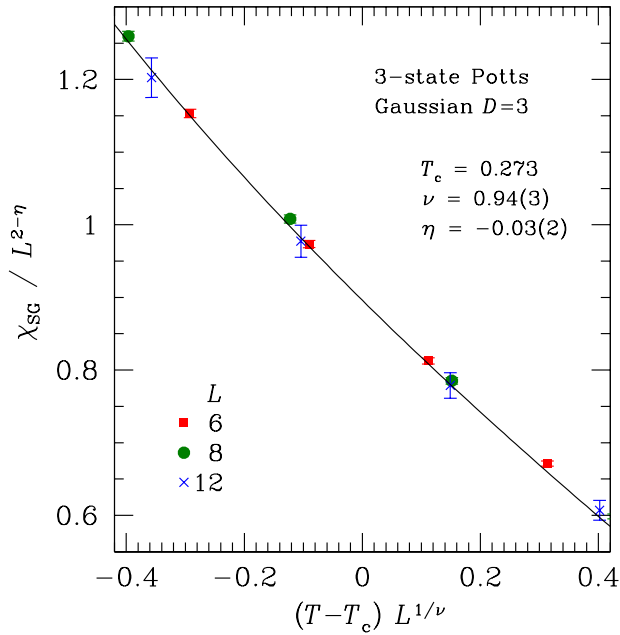}
\caption{(Color online) 
Scaling plot of $\chi_{\rm SG}/L^{2-\eta}$ vs $T$ for the 
three-state Potts glass with Gaussian interactions in three dimensions
according to Eq.~(\ref{eq:chisg_fss}) with $T_{\rm c}$ fixed to 0.273 
(determined from the
data for $\xi_L/L$ shown in Fig.~\ref{fig:xi_L_3state_gau_D3}). The fit to
Eq.~(\ref{eq:chisg_fss}) then gives $\eta = -0.03$ and $\nu = 0.94$.
}
\label{fig:3state_chiSG_Tc_fixed_from_xiL.inset}
\end{figure}
\end{center}

\begin{table}
\caption{Parameters of the simulation for the three-state Potts glass with Gaussian
interactions and $J_0 = 0$ in three dimensions. $N_{\sub{samp}}$ is the number of
samples.
$N_{\sub{sw}}$ is the total number of Monte Carlo sweeps used for measurement
for each of the
$2N_T$ replicas for a single sample. An equal number of sweeps is used for
equilibration; hence, the total number of MCS performed is $2 N_{\sub{sw}}$.
$T_{\sub{min}}$ is the lowest temperature
simulated, and $N_T$ is the number of temperatures used in the parallel
tempering method for each system size $L$.
\label{table:params_3state_Gau_3D}}
\begin{tabular*}{\columnwidth}{@{\extracolsep{\fill}}llllr}
\hline
\hline
$L$ & $T_{\mbox{\footnotesize min}}$ & $N_T$ & $N_{\mbox{\footnotesize samp}}$
& $N_{\mbox{\footnotesize sw}}$\\
\hline
4  & 0.2   & 18 & 9356 & 80000\\
6  & 0.2   & 18 & 3980 & 80000\\
8  & 0.2   & 20 & 4052 & 655360\\
12 & 0.248 & 28 & 352  & 16777216\\
\hline
\hline
\end{tabular*}
\end{table}

Next we discuss the results for $\chi_{\rm SG}$. We find that allowing 
$T_{\rm c}$ to
vary, in addition to $\nu$ and $\eta$, the value of $T_{\rm c}$ so obtained is
rather different from that obtained using data for $\xi_L/L$.
Since we believe the value of $T_{\rm c}$ obtained from $\xi_L/L$---namely,
$T_{\rm c} = 0.273$---to be our most accurate estimate, we
fix $T_{\rm c}$ to this value and fit $\eta$ and $\nu$ (and the polynomial
parameters $c_i$) to Eq.~(\ref{eq:chisg_fss}).
This gives $\nu = 0.94(3)$ and 
\begin{equation}
\eta=-0.03(2) \, .
\label{3d_gauss_eta}
\end{equation}
We show the corresponding plot in
Fig.~\ref{fig:3state_chiSG_Tc_fixed_from_xiL.inset}.  

If we analyze $\chi_{\rm SG}$ allowing $T_{\rm c}$ to vary, as well as the
exponents $\nu$ and $\eta$, we obtain
$T_{\rm c} = 0.254(23)$, $\nu = 0.99(8)$, and $\eta = -0.18(16)$. The transition
temperature and exponents are consistent, within the statistical
errors,
with those found for the same fit with $T_{\rm c}$ fixed to be 
$0.273$. We can also analyze the $\chi_{\rm SG}$ data by fixing \textit{both}
$T_{\rm c}$ and $\nu$ to the values obtained from $\xi_L/L$ and so obtain just
$\eta$. This gives $\eta = -0.06(2)$ which is consistent with the result in
Eq.~(\ref{3d_gauss_eta}).
As discussed in Sec.~\ref{sec:results_4d}, the results from $\xi_L/L$ are
expected to be more accurate than those from $\chi_{\rm SG}$.

The three-state Potts model is somewhat similar to the $XY$ spin glass in that the
spins point along directions in a two-dimensional plane. The difference is
that in the $XY$ model the spins can point in any direction in the plane while
in the three-state Potts model they can only point to one of three equally spaced
directions. In units where the standard deviation of $J'_{ij}$ is unity [see
Eq.~(\ref{simplexham})] a transition was found\cite{lee:03}
in the $XY$ spin glass at
$T_{\rm c}^{XY} = 0.34(2)$. Since our interactions $J_{ij}$ differ from the
$J'_{ij}$ by a factor of $(p-1)/p$ [see Eq.~(\ref{J'})] this corresponds to
$T_{\rm c}^{XY} = 0.23(2)$ in the units used here. We find a slightly larger
value for $T_{\rm c}$ in the Potts case, which is reasonable since
fluctuations in the Potts model
are presumably reduced relative to those in the $XY$ model
by the constraint on the spin directions. 

\begin{center}
\begin{figure}
\includegraphics[width=\figurewidth]{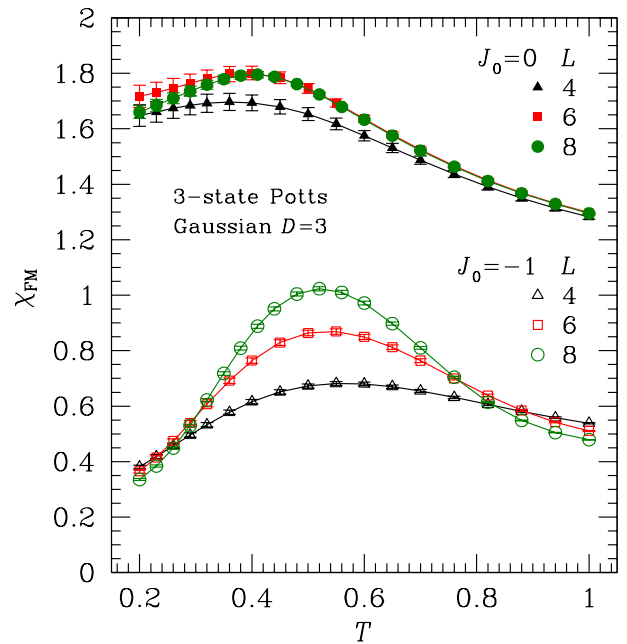}
\caption{(Color online) 
Ferromagnetic susceptibility $\chi_{\rm FM}$ against $T$ for the three-state 
Potts glass with Gaussian couplings
in three dimensions. The upper curves plotted with solid symbols correspond to
$J_0=0$ while the lower curves with open symbols correspond to $J_0=-1$,
where $J_0$ is the mean of the Gaussian interactions.
}
\label{fig:chiFM}
\end{figure}
\end{center}

\begin{table}
\caption{Parameters of the simulation for the three-state Potts glass with Gaussian
interactions in three dimensions used to compare results with
$J_0 = 0$ and $J_0 = -1$ where $J_0$ is the mean of the distribution.
See Table \ref{table:params_3state_Gau_3D} for details.
\label{table:params_3state_Gau_3D_J0}}
\begin{tabular*}{\columnwidth}{@{\extracolsep{\fill}}lllllr}
\hline
\hline
$L$ & $J_0$ & $T_{\mbox{\footnotesize min}}$ & $N_T$ & $N_{\mbox{\footnotesize samp}}$ & $N_{\mbox{\footnotesize sw}}$\\
\hline
4 & 0  & 0.2 & 18 & 1000 & 81920\\
6 & 0  & 0.2 & 18 & 1000 & 81920\\
8 & 0  & 0.2 & 20 & 2400 & 655360\\
\hline
4 & -1 & 0.2 & 18 & 1000 & 81920\\
6 & -1 & 0.2 & 18 & 1000 & 81920\\
8 & -1 & 0.2 & 20 & 2256 & 655360\\
\hline
\hline
\end{tabular*}
\end{table}

\begin{center}
\begin{figure}
\includegraphics[width=\figurewidth]{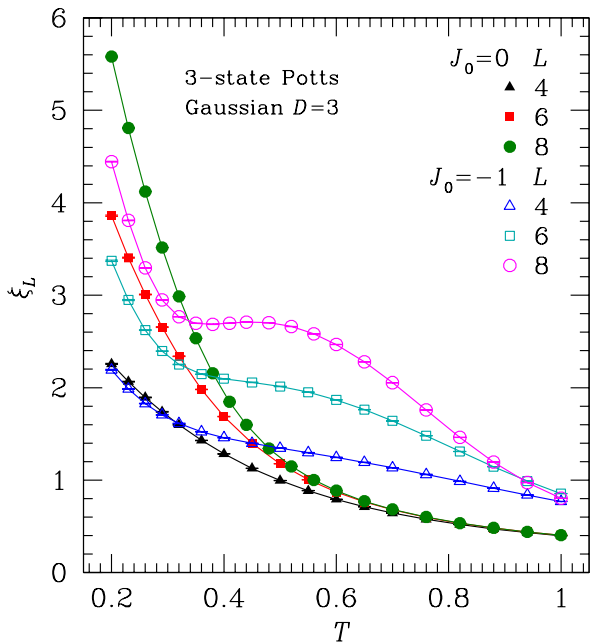}
\caption{(Color online) 
Graph of $\xi_L$ vs $T$ for the three-state Potts glass with Gaussian couplings
in three dimensions. The solid symbols correspond to $J_0 = 0$ while the open
symbols correspond to $J_0 = -1$.
}
\label{fig:xi_3state}
\end{figure}
\end{center}

\begin{center}
\begin{figure}
\includegraphics[width=\figurewidth]{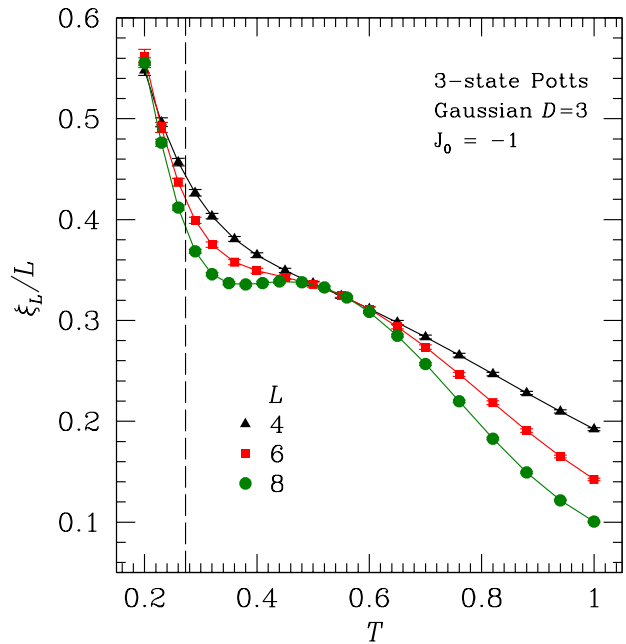}
\caption{(Color online) 
Graph of $\xi_L/L$ vs $T$ for the three-state Potts glass with Gaussian
couplings in three dimensions where the mean of the Gaussian distribution is
$J_0 = -1$. The vertical line is drawn at the transition temperature
$T_{\rm c} = 0.273$ of the $J_0 = 0$ case (see
Fig.~\ref{fig:xi_L_3state_gau_D3}).
}
\label{fig:xi_L_3state_mo}
\end{figure}
\end{center}

As mentioned before, the Potts glass with $p > 2$ develops ferromagnetic
correlations at low temperatures even with $J_0 = 0$. To see the extent of
this, we have performed additional simulations, with the parameters shown in
Table~\ref{table:params_3state_Gau_3D_J0}, for $J_0 = -1$ as well as $J_0 =
0$. We have calculated the ferromagnetic susceptibility, defined by
\begin{equation}
\chi_{\rm FM} = N[\langle \left|\mathbf{m}\right|^2 \rangle]_\av \, ,
\end{equation}
where $\mathbf{m} = N^{-1} \sum_i \mathbf{S}_i$,
and show the data in Fig.~\ref{fig:chiFM}.
For $J_0 = 0$, the ferromagnetic susceptibility grows slowly with temperature,
seems to be bounded from above, and only displays a small system-size 
dependence. We therefore think that ferromagnetic correlations should not 
seriously affect the results presented earlier in this subsection.
For $J_0 = -1$, the absolute value of $\chi_{\rm FM}$ is smaller, but shows
stronger size and temperature dependence.  It seems that $\chi_{\rm FM}$
initially increases as $T$ is reduced, but then is strongly suppressed as 
spin-glass correlations develop for $ T \to T_{\rm c} \simeq  0.27$.
The spin-glass correlation length for both
$J_0 = 0$ and $J_0 = -1$ is shown in Fig.~\ref{fig:xi_3state}.
The slope of the $J_0=-1$ data varies nonmonotonically with temperature,
reflecting the nonmonotonic behavior of the ferromagnetic 
correlations. By contrast, the $J_0=0$ data vary much more smoothly.
We further plot $\xi_L/L$ for $J_0 = -1$ in
Fig.~\ref{fig:xi_L_3state_mo}.
The observed nonmonotonic behavior of the slope suggests that corrections
to finite-size scaling will be large and that the data in
Fig.~\ref{fig:xi_L_3state_gau_D3} should
be more reliable in estimating $T_{\rm c}$. Nonetheless, the $J_0=-1$ data 
\textit{do} have
an intersection point, although at a lower temperature,
$T \simeq 0.20$, than the value of $0.273$ for the $J_0=0$ case.  We interpret
these data to indicate that $T_{\rm c} \simeq 0.20$ for the $J_0=-1$ model (of course
the value of $T_{\rm c}$ can vary with $J_0$).
However, data on a
larger range of sizes at lower temperature would be needed to confirm this.
Unfortunately,
equilibration times are very long at such low temperatures, so these
calculations were not feasible.

\begin{center}
\begin{figure}
\includegraphics[width=\figurewidth]{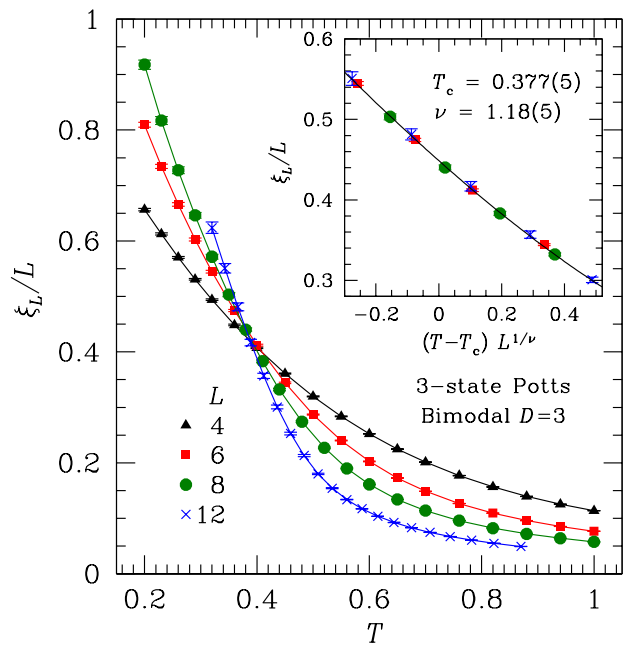}
\caption{(Color online) 
Graph of $\xi_L/L$ vs $T$ for the three-state Potts glass with a
bimodal coupling distribution
in three dimensions. The curves intersect at $T_{\rm c} \simeq 0.377$.
The inset shows a scaling plot according to Eq.~(\ref{eq:fss}) using
$T_{\rm c} = 0.377$ and $\nu = 1.18$.
}
\label{fig:xi_L_3state_pmJ_D3}
\end{figure}
\end{center}

\begin{center}
\begin{figure}
\includegraphics[width=\figurewidth]{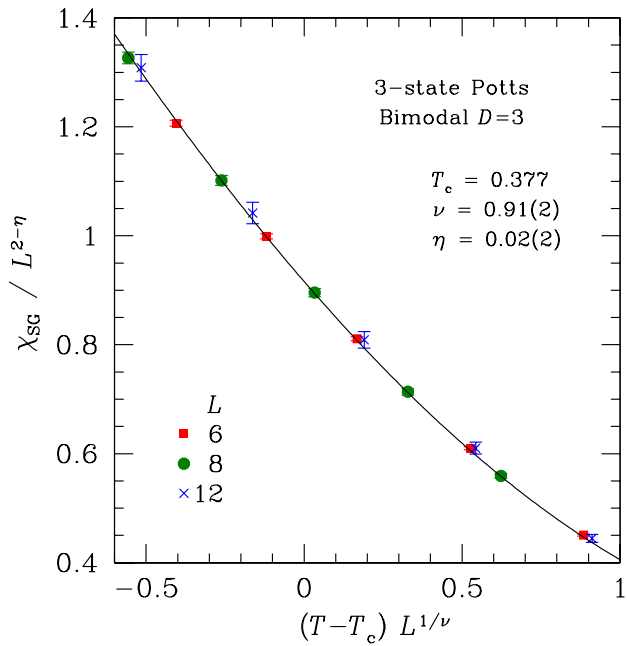}
\caption{(Color online) 
Scaling plot of $\chi_{\rm SG}$ according to
Eq.~(\ref{eq:chisg_fss}) with $T_{\rm c}$ fixed to 0.377 (determined from 
the data for $\xi_L/L$ shown in Fig.~\ref{fig:xi_L_3state_pmJ_D3})
for the three-state Potts glass with bimodal ($\pm J$) interactions in three
dimensions. The fit to
Eq.~(\ref{eq:chisg_fss}) then gives $\eta = 0.02$ and $\nu = 0.91$.
}
\label{fig:3state_pmJ_chiSG_Tc_fixed_from_xiL}
\end{figure}
\end{center}

\begin{table}
\caption{Parameters of the simulation for the three-state Potts glass with bimodal
interactions and $J_0 = 0$ in three dimensions. For further details see the
caption of Table \ref{table:params_3state_Gau_3D}.
\label{table:params_3state_pmJ_3D}}
\begin{tabular*}{\columnwidth}{@{\extracolsep{\fill}}llllr}
\hline
\hline
$L$ & $T_{\mbox{\footnotesize min}}$ & $N_T$ & $N_{\mbox{\footnotesize samp}}$ & $N_{\mbox{\footnotesize sw}}$\\
\hline
4  & 0.2   & 18 & 10000 & 5120\\
6  & 0.2   & 18 & 4971  & 40960\\
8  & 0.2   & 20 & 2046  & 1310720\\
12 & 0.320 & 20 & 550   & 4194304\\
\hline
\hline
\end{tabular*}
\end{table}

\subsection{Bimodal disorder ($\boldsymbol{d=3}$)}

We expect that the value of the lower critical dimension and other universal
quantities such as critical exponents are independent of the distribution
of the interactions. To verify this we present in this subsection data for
the three-state Potts glass in three dimensions with a bimodal ($\pm J$)
distribution for comparison with the above results for the Gaussian
distribution. The simulation parameters are shown in 
Table.~\ref{table:params_3state_pmJ_3D}.

Our results for $\xi_L/L$ are shown in 
Fig.~\ref{fig:xi_L_3state_pmJ_D3}. The data intersect at
$T_{\rm c} \simeq 0.38$, indicating a spin-glass transition.
Fitting the data for $\xi_L$ for $0.34 < T < 0.44$ and $L \ge 6$,
we obtain
\begin{equation}
T_{\rm c} = 0.377(5) \qquad \nu = 1.18(5) .
\label{res_3d_pmJ}
\end{equation}
The scaling plot according to Eq.~(\ref{eq:fss})
is shown in the inset of Fig.~\ref{fig:xi_L_3state_pmJ_D3}.

Similar to the data with Gaussian disorder,
the fits for $\chi_{\rm SG}$ give a different value for $T_{\rm c}$ 
from that obtained from data for $\xi_L/L$. Since we argue that the latter 
is more accurate, we have fitted
$\chi_{\rm SG}$ with $T_{\rm c}$ fixed to the value
$0.377$. This gives $\nu = 0.91(2)$ and
\begin{equation}
\eta = 0.02(2) \, .
\label{3d_pmJ_eta}
\end{equation}
These results are shown in Fig.~\ref{fig:3state_pmJ_chiSG_Tc_fixed_from_xiL}.

If we allow $T_{\rm c}$ to fluctuate, the fits to
the data for $\chi_{\rm SG}$ give $T_{\rm c} = 0.390(21)$,
$\nu = 0.87(7)$, and $\eta = 0.14(17)$. The error bars are larger than in the
analysis when $T_{\rm c}$ is fixed, but the results are consistent with 
the previous analysis. Note that for the Gaussian case, $T_{\rm c}$ estimated from
$\xi_L/L$ is greater than that from $\chi_{\rm SG}$, whereas it is lower for
the $\pm J$ distribution. However, these differences are within the
statistical errors and so are not significant.
As discussed in Sec.~\ref{sec:results_4d}, the results from $\xi_L/L$ are
expected to be more accurate than those from $\chi_{\rm SG}$.
We can also analyze the $\chi_{\rm SG}$ data by fixing \textit{both}
$T_{\rm c}$ and $\nu$ to the values obtained from $\xi_L/L$ and so obtain just
$\eta$. This gives $\eta = 0.03(2)$ which is consistent with the result in
Eq.~(\ref{3d_pmJ_eta}).

Comparing our results in this subsection with those above for the Gaussian
distribution, we see that in both cases we obtain
a finite $T_{\rm c}$, showing that $d_l < 3$.
This is in contrast to the $T = 0$ domain-wall renormalization group
calculations of Banavar and Cieplak\cite{banavar:89a,banavar:89b}
who find $d_l < 3$ for the Gaussian distribution
and $d_l > 3$ for the $\pm J$ distribution. One possible reason for this 
difference is the larger range of sizes that we are able to treat in 
the present study.

Furthermore, the values for 
the exponent $\nu$ agree within the statistical errors and those for
$\eta$ almost agree; see
Eqs.~(\ref{res_3d_gauss}), (\ref{3d_gauss_eta}), (\ref{res_3d_pmJ}), 
and (\ref{3d_pmJ_eta}). The small additional difference in the the values for
$\eta$ can be attributed
to systematic corrections to scaling. Hence our results
are consistent with universality.

\subsection{Gaussian disorder ($\boldsymbol{d=4}$)}
\label{sec:results_4d}

\begin{table}
\caption{Parameters of the simulation for the three-state Potts glass with Gaussian
interactions and $J_0 = 0$ in four dimensions. For further details see the caption
of Table \ref{table:params_3state_Gau_3D}.
\label{table:params_3state_Gau_4D}}
\begin{tabular*}{\columnwidth}{@{\extracolsep{\fill}}llllr}
\hline
\hline
$L$ & $T_{\mbox{\footnotesize min}}$ & $N_T$ & $N_{\mbox{\footnotesize samp}}$ & $N_{\mbox{\footnotesize sw}}$\\
\hline
3 & 0.2   & 18 & 1044 & 40960\\
4 & 0.2   & 18 & 2691 & 40960\\
5 & 0.350 & 17 & 1000 & 524288\\
6 & 0.408 & 18 & 700  & 327680\\
8 & 0.490 & 22 & 623  & 524288\\
\hline
\hline
\end{tabular*}
\end{table}

\begin{center}
\begin{figure}
\includegraphics[width=\figurewidth]{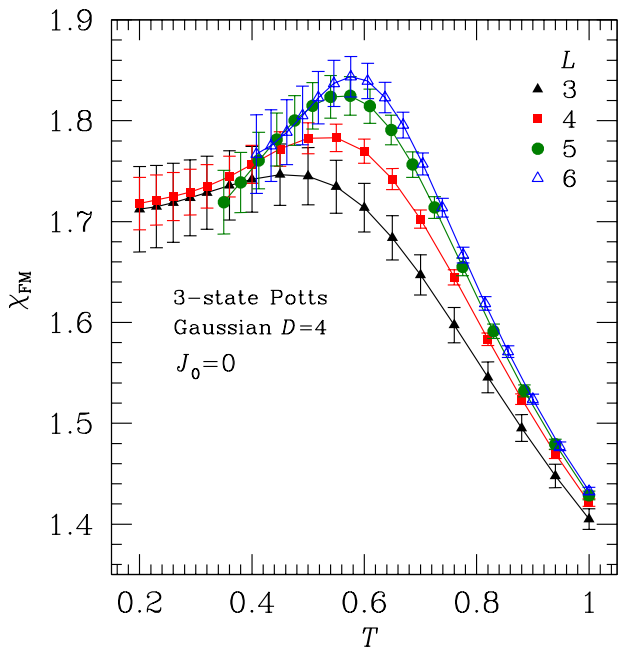}
\caption{(Color online) 
Graph of $\chi_{\rm FM}$ vs $T$ for the three-state Potts glass
with Gaussian couplings with $J_0 = 0$ in four dimensions.
}
\label{fig:chiFM_3state_4D}
\end{figure}
\end{center}

For the three-state Potts glass in four dimensions we first show
data for $\chi_{\rm FM}$ with $J_0 = 0$ in Fig.~\ref{fig:chiFM_3state_4D}
to see whether significant ferromagnetic correlations develop.
The behavior is very similar to the case in $d = 3$ (see
Fig.~\ref{fig:chiFM}) since $\chi_{\rm FM}$ remains
small and relatively size independent. This indicates that ferromagnetic
couplings, while present, are relatively weak thus we set $J_0 = 0$ in the
simulations of the three-state model in $d=4$.
The simulation parameters
are presented in Table~\ref{table:params_3state_Gau_4D}.

\begin{center}
\begin{figure}
\includegraphics[width=\figurewidth]{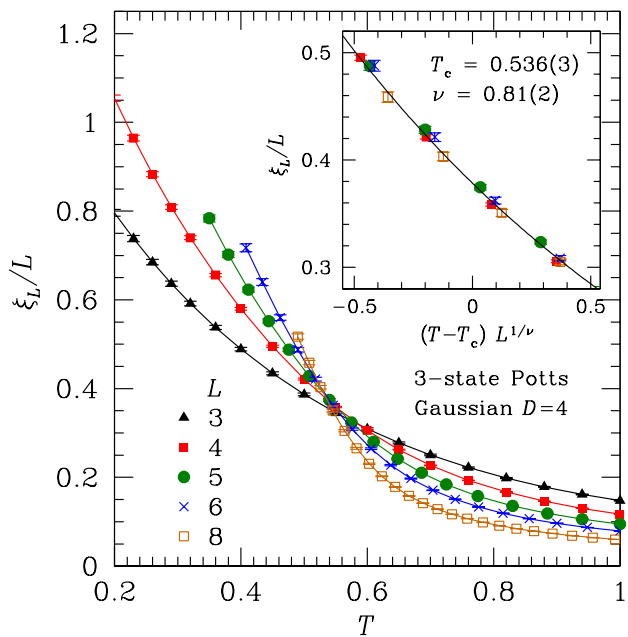}
\caption{(Color online) 
Correlation length $\xi_L/L$ vs $T$ for the three-state Potts glass
with Gaussian couplings
in four dimensions. The curves intersect at $T_{\rm c} \simeq 0.54$.
The inset shows a scaling plot according to Eq.~(\ref{eq:fss}) using
$T_{\rm c} = 0.536$ and $\nu = 0.81$.
}
\label{fig:xi_4D}
\end{figure}
\end{center}

Next we plot $\xi_L/L$ in Fig.~\ref{fig:xi_4D}, where the data intersect 
at  $T_{\rm c} \approx 0.54$. Fitting the data for $\xi_L$ for $0.49 < T
< 0.59$ and $L \ge 4$, we obtain the best fit as
\begin{equation}
T_{\rm c} = 0.536(3), \qquad \nu = 0.81(2) \, . 
\label{4d:xi_L}
\end{equation}
The scaling plot according to Eq.~(\ref{eq:fss}) is shown in the inset of
Fig.~\ref{fig:xi_4D}.

Our value for $T_{\rm c}$ is very different from that of Scheucher and
Reger\cite{scheucher:93} who estimate $T_{\rm c} = 0.25(5)$. However, in the
vicinity of their estimate for $T_{\rm c}$ they could equilibrate only 
two small sizes $L=3$ and 4. Furthermore, they used the Binder
ratio\cite{binder:81} (a ratio of moments of the
order parameter) which even fails to intersect\cite{dillmann:98} at the known
transition temperature in the mean-field Potts model. Hence we feel that our
approach, using the correlation length, is more reliable, especially since we
are able to use parallel tempering to simulate larger sizes at lower $T$ than
was possible for Scheucher and Reger in the ``pre-parallel tempering days.''

\begin{center}
\begin{figure}
\includegraphics[width=\figurewidth]{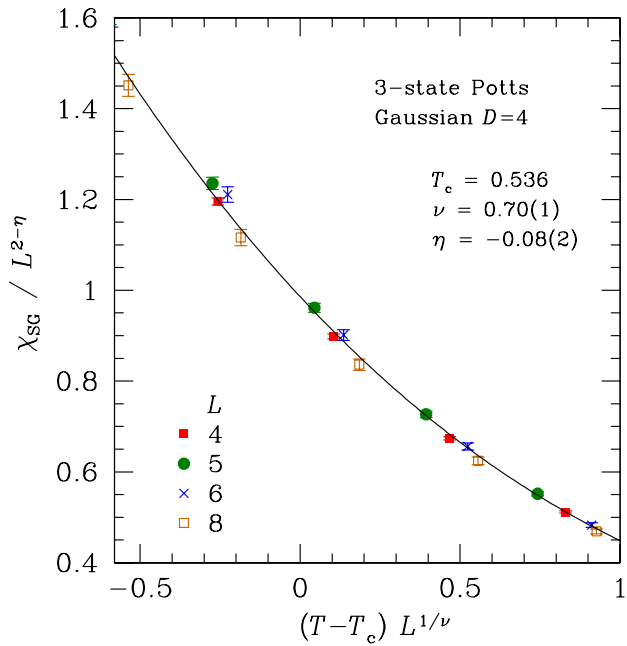}
\caption{(Color online) 
Scaling plot of $\chi_{\rm SG}$ for the three-state Potts glass
with Gaussian interactions in four dimensions according to
Eq.~(\ref{eq:chisg_fss}) with $T_{\rm c}$ fixed to the value
$0.536$, which is obtained from the data for $\xi_L/L$ shown 
in Fig.~\ref{fig:xi_4D}. The fit to Eq.~(\ref{eq:chisg_fss}) then gives
$\eta = -0.08$ and $\nu = 0.70$.
}
\label{fig:4D_chiSG_Tc_fixed_from_xiL}
\end{figure}
\end{center}

We next discuss the data for $\chi_{\rm SG}$.
First we fix $T_{\rm c}$ to be the value 0.536 obtained from $\xi_L/L$.
Fitting to Eq.~(\ref{eq:chisg_fss})
then gives $\nu = 0.70(1)$ and
\begin{equation}
\label{4d_eta_nu}
\eta = -0.08(2) .
\end{equation}
This is shown in Fig.~\ref{fig:4D_chiSG_Tc_fixed_from_xiL}.
We can also analyze the $\chi_{\rm SG}$ data by fixing \textit{both}
$T_{\rm c}$ and $\nu$ to the values obtained from $\xi_L/L$, and so obtain just
$\eta$. This gives $\eta = -0.08(2)$ which agrees with the result in
Eq.~(\ref{4d_eta_nu}).

However, if we allow $T_{\rm c}$ to vary, as well as $\nu$ and $\eta$,
we obtain
$T_{\rm c} = 0.505(14)$, $\nu = 0.75(3)$, and $\eta = -0.46(16)$. 
We see that $T_{\rm c}$
is substantially lower than in Eq.~(\ref{4d:xi_L}). The resulting value of
$\nu$ is not very different than that obtained in
Eq.~(\ref{4d:xi_L}) but the value of $\eta$ is \textit{significantly
more negative}
than that obtained in Eq.~(\ref{4d_eta_nu}). These differences are due to
corrections to finite-size scaling.
A related problem of different exponents from $\chi_{\rm SG}$ and $\xi_L/L$
was also found\cite{katzgraber:06} for the Ising spin glass (although there the
main discrepancy involved $\nu$ rather than $\eta$). They argued that
the data for
$\xi_L/L$ is
the most accurate way to determine $T_{\rm c}$ and $\nu$.
Since it is dimensionless, the
data give clean intersections without having to divide by an unknown
power of $L$. Reference~\onlinecite{katzgraber:06} confirmed
this supposition by
reanalyzing their data in an alternative way proposed in
Ref.~\onlinecite{campbell:06}. The results from $\xi_L/L$ did not change
significantly, but the value for $\nu$ from $\chi_{\rm SG}$
changed considerably and became much
closer to the value found from $\xi_L/L$.

Hence, for the present case,
we feel that the values of $\nu$ in Eq.~(\ref{4d:xi_L})
(obtained from $\xi_L/L$) and $\eta$ in Eq.~(\ref{4d_eta_nu})
(obtained from $\chi_{\rm SG}$ with $T_{\rm c}$ fixed at the value from  
$\xi_L/L$) are
the most accurate ones. An estimate for the order parameter exponent $\beta$
is then obtained from the scaling relation
\begin{equation}
\beta = (d - 2 + \eta) \, \frac{\nu}{2} \, ,
\end{equation}
which gives 
\begin{equation}
\beta =  0.78(3) \, .
\end{equation}
This is very different from the value $\beta \simeq 0$ obtained by Scheucher
and Reger.\cite{scheucher:93} Therefore, we find that
the transition in the three-state Potts
glass in four dimensions is actually not in the mean-field universality 
class\cite{gross:85} for $p > 4$.

\section{Ten-state Potts glass in $\boldsymbol{d=3}$ (Gaussian disorder)}
\label{sec:results_10state}

\begin{center}
\begin{figure}
\includegraphics[width=\figurewidth]{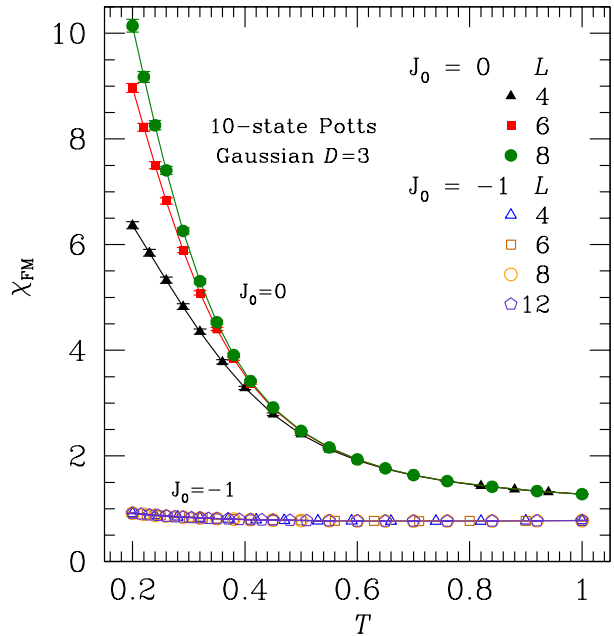}
\caption{(Color online) 
Ferromagnetic susceptibility $\chi_{\rm FM}$ vs $T$ for the ten-state 
Potts glass with Gaussian couplings in three dimensions. For $J_0 = -1$, 
represented by the open symbols, $\chi_{\rm FM}$ remains small and 
size independent. In contrast, $\chi_{\rm FM}$ for $J_0 = 0$ (solid symbols) 
becomes very large and shows substantial finite-size effects.
}
\label{fig:chiFM_10state}
\end{figure}
\end{center}

\begin{center}
\begin{figure}
\includegraphics[width=\figurewidth]{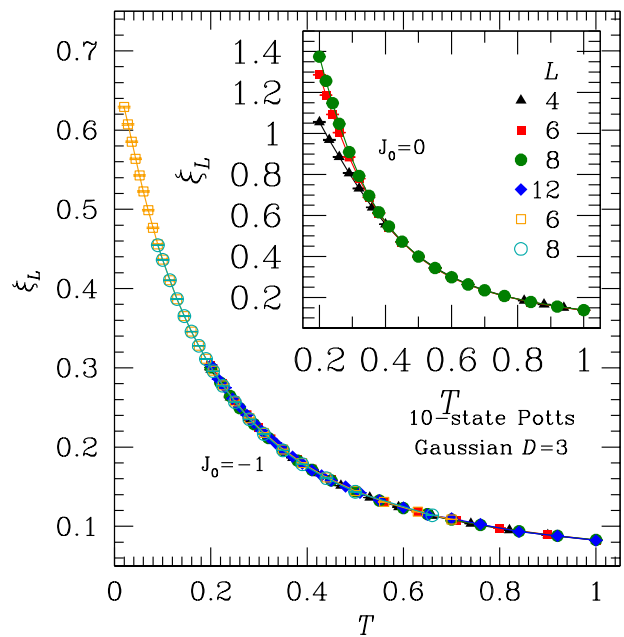}
\caption{(Color online) 
Correlation length $\xi_L$ (not $\xi_L/L$)
against $T$ for the ten-state Potts glass in three dimensions. The main
panel shows the case for $J_0 = -1$ where there are hardly any finite-size
effects. The inset shows the same for $J_0 = 0$ showing the 
size dependence of $\xi_L$ in this case.
}
\label{fig:xi_10state}
\end{figure}
\end{center}

Finally, we turn our attention to the ten-state Potts glass in three dimensions
with Gaussian interactions. We set the mean of the Gaussian interactions to
$J_0 = -1$ in order to suppress ferromagnetic ordering.\cite{brangian:03} The
effectiveness of this is shown in Fig.~\ref{fig:chiFM_10state}, where the
ferromagnetic susceptibility $\chi_{\rm FM}$ is seen to be very small and does
not exhibit measurable finite-size effects. However, if we take $J_0 = 0$,
which is also shown, $\chi_{\rm FM}$ becomes large and size dependent,
indicating that ferromagnetic ordering probably takes place.  The simulation
parameters used for the ten-state Potts glass for $J_0=-1$ are shown in
Table~\ref{table:params_10state}.  The second set of parameters for $L=6$ and
$8$ is for additional runs to probe the very-low-temperature behavior.

A plot of $\xi_L$ (\textit{not} divided by $L$) against temperature for $J_0 =
-1$ is shown in Fig.~\ref{fig:xi_10state}. The correlation length shows no
size dependence and stays very small, less than one lattice spacing, even at
extremely low temperatures. We conclude, in agreement with Brangian {\it et
al.},\cite{brangian:03} that there is clearly no spin-glass transition in this
model.

\begin{table}
\caption{Parameters of the simulation for the ten-state Potts
glass with Gaussian
interactions with
$J_0=-1$ in three dimensions.
The second set of parameters for $L = 6$ and $8$ is for additional runs to
probe the very low-temperature behavior of the model.
\label{table:params_10state}}
\begin{tabular*}{\columnwidth}{@{\extracolsep{\fill}}llllr}
\hline
\hline
$L$ & $T_{\mbox{\footnotesize min}}$ & $N_T$ & $N_{\mbox{\footnotesize samp}}$ & $N_{\mbox{\footnotesize sw}}$\\
\hline
4  & 0.2 & 14 & 1000 & 81920\\
6  & 0.2 & 17 & 1000 & 131072\\
8  & 0.2 & 19 & 998  & 262144\\
12 & 0.2 & 26 & 343  & 20480\\
\hline
6 & 0.02 & 28 & 469  & 262144\\
8 & 0.08 & 18 & 600  & 131072\\
\hline
\hline
\end{tabular*}
\end{table}

Since the nearest-neighbor ten-state Potts glass does not even come close to
having a transition down to $T=0$, its behavior is very different from the
mean-field (infinite-range) case, which has two transitions. Although the
mean-field version provides a plausible model for supercooled liquids as 
they are
cooled down towards the glass transition, the nearest-neighbor model does not.
It might therefore be interesting to investigate a Potts glass that has
somewhat longer-range interactions\cite{brangian:03} to see if its behavior
corresponds more closely to that of supercooled liquids. 

\section{Conclusions}
\label{sec:conclusions}

\begin{table}[ht]
\caption{
Summary of results for
the four different $p$-state Potts glasses studied. 
The values for $T_{\rm c}$ and $\nu$
are obtained
from analyzing data for $\xi_L/L$, which we argue give the most reliable
results. The values for $\eta$ are obtained from data for $\chi_{\rm SG}$ with
only $T_{\rm c}$ fixed to its value obtained from $\xi_L$.
\label{table:summary}
}
\begin{tabular*}{\columnwidth}{@{\extracolsep{\fill}} l l c c c r}
\hline
\hline
$p$ & Disorder & $d$ & $T_{\rm c}$ & $\nu$ & $\eta$\\
\hline
3 & Gaussian  & 3 & $0.273(5)$ & $1.30(7)$ & $-0.03(2)$ \\
3 & Bimodal ($\pm J)$   & 3 & $0.377(5)$ & $1.18(5)$ & $ 0.02(2)$ \\
3 & Gaussian  & 4 & $0.536(3)$ & $0.81(2)$ & $-0.08(2)$ \\
10& Gaussian  & 3 & \multicolumn{3}{c}{No transition}\\
\hline
\hline
\end{tabular*}
\end{table}

We show a summary of our results for the various models in 
Table.~\ref{table:summary}.
We find strong evidence for a finite transition temperature $T_{\rm c}$ in the
three-state Potts glass in three dimensions, with universal critical exponents.
This differs from several earlier studies which claimed that the lower
critical dimension is equal or close to $3$. In
four dimensions, the three-state Potts glass has a transition which is quite
different from that proposed by
Scheucher and Reger\cite{scheucher:93}---namely
the discontinuous transition found in mean-field theory for $p > 4$.
The ten-state
Potts glass in three dimensions, which in its mean-field incarnation is 
often used to describe structural glasses, is very far from having a phase
transition into a glassy phase at any temperature since its
correlation length remains very small even down to extremely low temperatures,
in agreement with the work of Brangian {\it et al.}\cite{brangian:03}

\begin{acknowledgments}
The simulations were performed in part on the Hreidar and Gonzales clusters
at ETH Z\"urich.
L.W.L.~and A.P.Y.~acknowledge support from the National Science Foundation 
under Grant No.~DMR 0337049. We also acknowledge a generous provision of
computer time on a G5 cluster from the Hierarchical Systems Research
Foundation.
\end{acknowledgments}

\bibliography{refs}

\end{document}